# Noise effect on 2D photoluminescence decay analysis using the RATS method in single-pixel camera configuration

**JIŘÍ JUNEK,**[1,2,*] **KAREL ŽÍDEK,**[1]

[1]*Regional Center for Special Optics and Optoelectronic Systems (TOPTEC), Institute of Plasma Physics, Czech Academy of Science v.v.i., Za Slovankou 1782/3, 182 00 Prague 8, Czech Republic*
[2]*Technical University in Liberec, Faculty of Mechatronics, Informatics and Interdisciplinary Studies, Studentská 1402/2, 461 17 Liberec, Czech Republic*
*junekj@ipp.cas.cz*

**Abstract:** Using a random temporal signal for sample excitation (RATS method) is a new, efficient approach to measuring photoluminescence (PL) dynamics. The method can be used in single-point measurement (0D), but also it can be converted to PL decay imaging (2D) using a single-pixel camera configuration. In both cases, the reconstruction of the PL decay and PL snapshot is affected by ubiquitous noise**.** This article provides a detailed analysis of the noise effect on the RATS method and possible strategies for its elimination. We carried out an extensive set of simulations focusing on the effect of noise introduced through the random excitation signal and the corresponding PL waveform. We show that the PL signal noise level is critical for the method. Furthermore, we analyze the role of acquisition time, where we demonstrate the need for a non-periodic excitation signal. We show that it is beneficial to increase the acquisition time while increasing the number of measurements in the single-pixel camera configuration has a minimal effect above a certain threshold. Finally, we study the effect of a regularization parameter used in the deconvolution step, and we observe that there is an optimum value set by the noise present in the PL dataset. Our results provide a guideline for optimization of the RATS measurement, but also study effects generally occurring in PL decay measurements methods relying on the deconvolution step.

## 1. Introduction

Fluorescence lifetime imaging (FLIM) is a frequent spectrometric analysis in biology [1,2], chemistry [3], and materials engineering [4]. There are various techniques for measuring or evaluating PL dynamics like time-correlated single photon counting [5], gated PL counting [6], streak camera [7], time-domain or frequency-domain analog recording technique [8,9]. Another available method for PL decay measurement is based on random excitation of the measured sample - the RATS method [10], which could also be set in single-pixel camera (SPC) configuration [11] and carry out the FLIM experiment. We abbreviate this measurement as 2D-RATS.

The 2D-RATS method combines the SPC technique with the excitation of the PL with random patterns (masks) blinking randomly in time. Each mask leads to a specific PL decay, which is measured for a set of uncorrelated masks. Then, using algorithms for undetermined systems, it is possible to reconstruct PL decays in each individual pixel of the 2D scene. The quality of reconstruction is affected by noise, which is inevitably present in all measured data. Therefore, the noise analysis of the method is an essential piece of knowledge that allows optimization of the method and identifies the most sensitive aspects of the setup. Moreover, many results valid for the RATS method are also relevant for the PL measurement, where the deconvolution step is used [12].

In this article, we carry out a detailed study of various noise effects on the 2D-RATS measurement and their impact on the resulting FLIM information. Our analysis is based on extensive simulations and a set of synthetic data faithfully following the real experimental conditions for the SPC experiment carried with a digital micro-mirror device. We analyze the benefits of various approaches to noise reduction. Besides the generally used option to prolong the acquisition time [13], the deconvolution step in the RATS method allows us to optimize the regularization parameter of the deconvolution [14,15]. Due to the random nature of the excitation signal in RATS, we avoid the use of any mathematical filters which can be apparently optimized for one type of signal, but due to the possibility of choosing a different signal frequency and sampling, such a filter can be difficult to transfer to a general case.

Our article provides guidelines for optimizing experiments for PL decay measurement and FLIM, where the deconvolution-based retrieval of PL dynamics is used. We demonstrate that the noise present in the PL dataset is the most critical factor. At the same time, we show that the resulting noise level in the FLIM dataset can be highly decreased by using a longer non-periodic excitation signal or by optimizing the regularization parameter. The parameter features an optimum value for a given noise level. We point out the issues connected to a periodic excitation signal.

In particular, we demonstrate that with proper choice of regularization parameter and acquisition time, the RATS method can attain results that are not distorted at all and can accurately map a 2D scene even with a relatively high noise level (3%).

## 2. RATS method

We will first introduce the principles of the single-spot 0D-RATS method and imaging 2D-RATS measurement. The cornerstone of the RATS method is sample excitation with a random signal $I_{EXC}$. PL signal $I_{PL}$ is a convolution of $I_{EXC}$ and PL decay $I_D$ (Eq.(1)).

$$I_{PL} = I_{EXC} * I_D. \qquad (1)$$

This fact can be used to recover the PL decay from $I_{PL}$ by using the Fourier transform and the convolution theorem. The deconvolution is attainable only for the frequencies where the excitation signal has a non-zero amplitude. Nevertheless, due to random excitation, a single measurement contains a broad range of frequencies in a single dataset. PL decay $I_D$ could then be determined via deconvolution (Eq.(2)).

$$I_D = Re\left\{ \mathbb{F}^{-1}\left[ \frac{\mathbb{F}(I_{PL})\,\mathbb{F}^*(I_{EXC})}{\mathbb{F}(I_{EXC})\mathbb{F}^*(I_{EXC}) + \varepsilon\overline{\mathbb{F}(I_{EXC})\mathbb{F}^*(I_{EXC})}} \right] \right\}. \qquad (2)$$

The 2D-RATS measurement is carried out in an SPC configuration, as we described in our previous article [11]. We point the reader to numerous articles summarizing the SPC experiment for more details [16-19]. In our implementation of the SPC experiment, the sample was illuminated with a set of random excitation patterns (masks). The excitation masks were "blinking" with the same waveform $I_{EXC}(t)$, and the whole PL from each illuminated point $i$ is collected to a single-pixel detector. Hence, the photoluminescence $I_{PL}$ is given as a summation of $I_{PL}(i)$ signals. Then Eq.(1) can be rewritten for the total PL intensity as:

$$I_{PL} = \sum_{i=1}^{n} I_{PL}(i) = I_{EXC} * \sum_{i=1}^{n} I_D(i). \qquad (3)$$

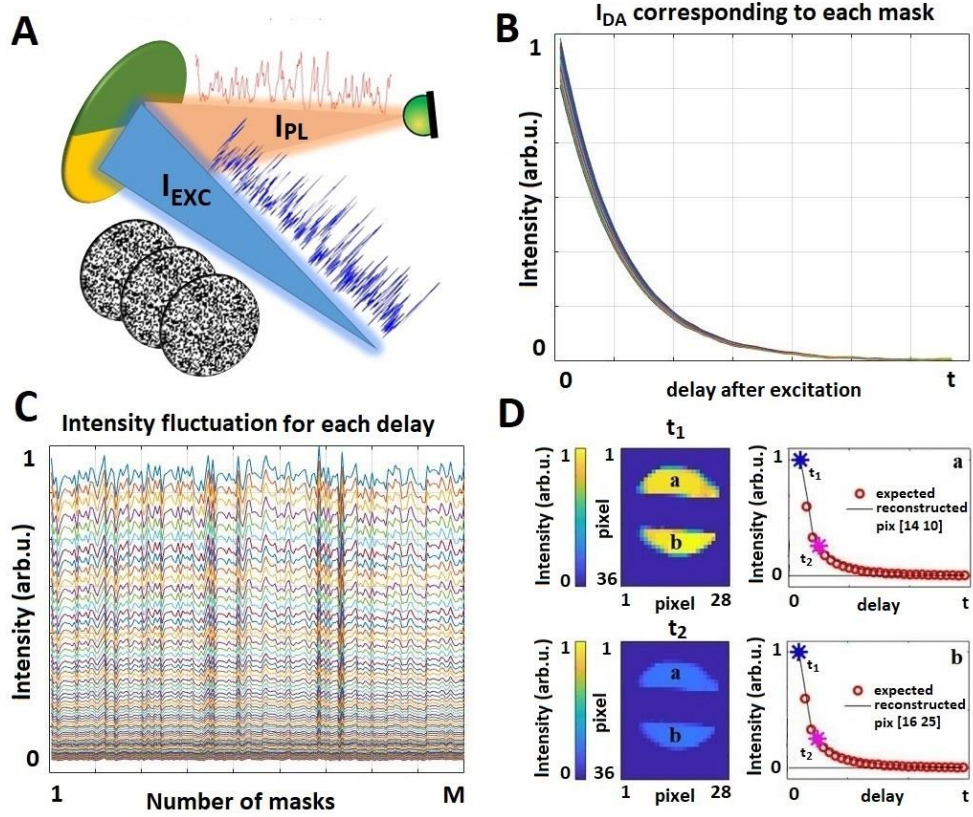

Figure 1: (A) Scheme of 2D-RATS approach, where the sample is illuminated with a set of random patterns (masks) blinking according to $I_{EXC}$. The total PL intensity waveform $I_{PL}$ is detected for each mask with a single-pixel detector. (B) Example of a set of calculated $I_{DA}$'s for corresponding masks – see Eq.(4). (C) Intensity fluctuation for each delay after excitation of $I_{DA}$. Knowing the fluctuation ($I_{SPC}$ signal) and set of masks makes it possible to determine the PL map m at a given delay after excitation of $I_{DA}$ using compressed sensing algorithms. (D) An example of reconstructed PL maps in delays $t_1$ and $t_2$. Plotted $I_D$'s corresponding to pixel [14 10] (area "a") and to pixel [16 25] (area "b").

The number of excitation masks $M$ is set by the number of image pixels $N$ and the so-called compression ratio $k = M/N$. Because masks are not coherent, i.e., one pattern is not correlated with each other, each mask illuminates a different combination of the sample pixels, and, therefore, each mask represents its own $I_{PL}$, which can be used to retrieve the PL decay $I_{DA}$:

$$I_{DA} = Re\left\{ \mathbb{F}^{-1}\left[ \frac{\mathbb{F}\left(\sum_{i=1}^{n} I_{PL(i)}\right) \mathbb{F}^{*}\left(I_{Exc}\right)}{\mathbb{F}\left(I_{Exc}\right)\mathbb{F}^{*}\left(I_{Exc}\right) + \varepsilon \overline{\mathbb{F}\left(I_{Exc}\right)\mathbb{F}^{*}\left(I_{Exc}\right)}} \right] \right\}. \qquad (4)$$

PL snapshot $m(t)$ can be reconstructed in each delay after excitation $t$ using the calculated $I_{DA}(t)$ curves. For the given delay and each excitation mask, the $I_{DA}(t)$ provides a dataset $I_{SPC}$ describing intensity fluctuation of $M$ values – see Fig. 1(C), where each curve corresponds to a single $I_{SPC}$ dataset. Then PL snapshot m$(t)$ is then retrieved from the underdetermined dataset by using a standard SPC retrieval:

$$min\{\|Am(t) - I_{SPC}\|_2^2 + TV(m(t))\}. \tag{5}$$

The sensing matrix $A$ is formed from vectorized excitation patterns, TV stands for the total variation calculation.

By reconstructing a temporal snapshot for each delay $t$, we get a 3D datacube, which contains PL decay in every $i$-th pixel of the sample $I_D(i,t)$. The whole concept is illustrated and summarized in Fig. 1.

## 3. Noise effect simulation

A small contribution of noise can be expected in each measured signal in the experiment. In order to maintain stable conditions, the role of the noise was explored through simulations. A primary signal $I_{EXC}$ was simulated using temporal speckles patterns [20], which are cornerstones of a random analog signal generator presented in our previous work [10]. The length of the simulated $I_{EXC}$ signal was 0.1 s with an impulse response function of FWHM of 2.07μs. The PL decay $I_D$ was considered with the lifetime $\tau = 20$ μs. Used noise levels throughout the article are consistent with the noise that can be expected in a realistic experiment [12]. The excitation signal $I_{EXC}$ was simulated as non-periodic. The importance of a non-periodic signal will be explained in section 4.1. It is worth noting that with the exception of section 4.3, the regularization parameter $\varepsilon$ (Eq.(2) and Eq.(4), respectively) is considered as $\varepsilon = 0.1$.

### 3.1 0D-RATS

To demonstrate the basic steps, we shall start with the non-imaging RATS, i.e., 0D-RATS. We use two datasets of $I_{EXC}$ and $I_{PL}$, which are subsequently used to retrieve the PL decay via Eq.(2). Therefore, we initially simulated the effect of noise present in these two datasets.

Firstly, we added white noise to $I_{EXC}$, while $I_{PL}$ was kept absolutely noiseless (Fig. 2(A)) and vice versa (Fig. 2(B)). In both datasets in Fig. 2, the signal-to-noise ratio (SNR) was 15.2 dB which corresponds to 3% percent of noise in the system (see Eq.(6)).

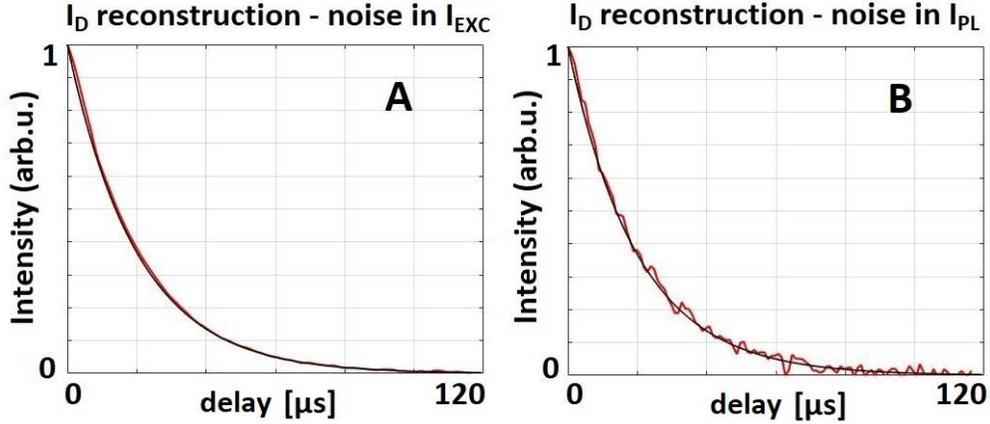

Fig 2: (A) Reconstructed $I_D$ with a noise level of 3% in $I_{EXC}$, which corresponds with SNR about 15.2 dB. $I_{PL}$ was assumed noiseless. (B) Reconstructed $I_D$ with a noise level of 3% in $I_{PL}$, corresponding with SNR about 15.2 dB. $I_{EXC}$ was assumed noiseless.

$$SNR[dB] = -10\log_{10}\left(\frac{noise[\%]}{100}\right). \tag{6}$$

For both retrieved decays in Fig. 2 (red lines), the noise had an effect on reconstructed $I_D$, but the measured PL dynamics were not biased, i.e., the overall decay shape was not altered, and it was in perfect agreement with the expected output (black lines). Nevertheless, we observed that the noise in the $I_{PL}$ dataset had a significantly more pronounced effect on the resulting noise in the retrieved PL decay (Fig. 2(B)).

We can quantify the noise level by the root-mean-square error, which reaches $RMSE_{PL} = 12,9.10^{-3}$ for the noise in the PL signal and $RMSE_{EXC} = 2,8.10^{-3}$ for the noise in the excitation signal. That suggests that regularization in the denominator of Eq.(2) has a more pronounced effect on $I_{EXC}$ than $I_{PL}$, which becomes the dominating source of noise in the retrieved decay.

### 3.2 2D-RATS

As the next step, we also simulated the effect of noise in the $I_{PL}$ and $I_{EXC}$ signal on the 2D-RATS experiment. Here, the situation is more complex because the noise present in the $I_{PL}$ and $I_{EXC}$ datasets is first transposed into the noise of the retrieved PL decay curve $I_{DA}(t)$ from Eq.(4). The noise present in these curves is then propagating into the SPC signal $I_{SPC}$, which is then used for the retrieval of the set of PL snapshots for each delay in Eq.(5).

All simulated reconstructions were performed using binary masks. We used the reconstruction algorithm based on the sparsity of total variance TVAL3 [21,22]. The main parameters of the TVAL3 algorithm were: *mu* ($2^{11}$), *beta* ($2^{7}$). As a measured sample, we used Matlab predefined image Phantom, where the PL decay with a single lifetime of $\tau = 20$ μs was set to be constant all over the Phantom's "body". Acquisition time and other conditions simulating real experiments were kept the same as for the 0D-RATS simulations above.

We carried out the same calculation as we presented in Section 3.1, and we show in Fig. 3 the retrieved PL snapshots $m(t)$ in the peak PL intensity, i.e., $t = 0$ μs, where the contrast of the PL signal is the highest. The case was studied for three different SPC compression ratios $k = 0.4, 0.6, 0.8$ and four signal-to-noise ratios $SNR = 15.2, 18.2, 20, 23$ dB, which corresponds to 3%, 1.5%, 1%, 0.5% level of noise in the signal.

The results are summarized in Fig. 3 for the noise introduced in the $I_{EXC}$ signal (left-hand side) and the noise introduced in the $I_{PL}$ signal (right-hand side), and the corresponding resulting noise levels are presented in Table 1.

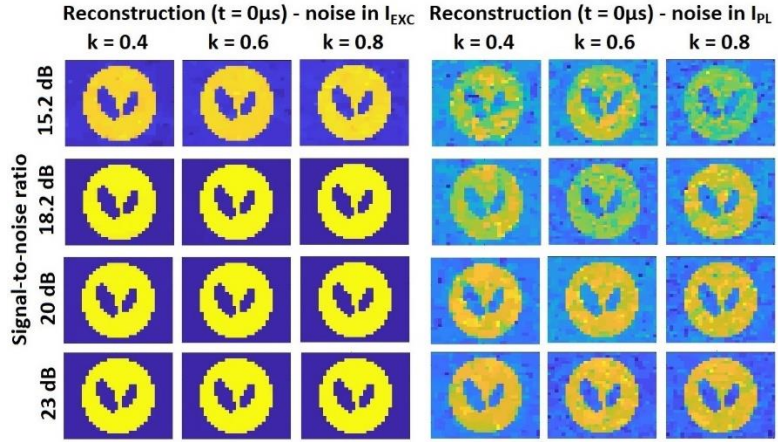

Figure 3: Left part: reconstructions of PL map m in delay with maximal intensity (t = 0 μs) in case of noise in $I_{EXC}$. Right part: reconstructions of PL map m in delay with maximal intensity (t = 0 μs) in case of noise in the $I_{PL}$. The noise parameters are for both SNR = 15.2, 18.2, 20, 25 dB (rows) and the compression ratios are k = 0.4, 0.6, 0.8 (columns).

Analogously to the 0D-RATS, the noise in $I_{PL}$ has a significantly worse effect on the image retrieval than the noise in the $I_{EXC}$ signal. By comparing different noise levels (compare lines in Fig. 3), one can see that the *SNR* of the $I_{PL}$ is the main limiting factor in the snapshot retrieval. The effect of compression ratio (compare columns in Fig. 3) does not play any significant role for the values above 0.4. This means that it is not efficient to compensate for higher noise in the measured signal by simply increasing the number of measured excitation masks.

To quantify the effect of each parameter, we introduced three different measures. Firstly, we can judge the image reconstruction quality, where we compare the retrieved snapshot $m$ (in delay $t = 0$ µs) with the actual image used in simulations $U$ by using Frobenius norm of the two:

$$R = \frac{\|m - U\|_F}{\|U\|_F}. \tag{7}$$

We can also focus on the PL dynamics for each pixel $i$, where the reconstructed decay curve $I_{DREC}(i)$ is normalized and compared with the actual decay curve $I_D(i)$, which was used to simulate the data. The comparison was made from $t_0 = 0$ µs to $t_\tau = 120$ µs, which is a sufficient temporal range for the decay curve with a lifetime of 20 µs. We denote this error as the decay deviation $\sigma$:

$$\sigma = \frac{1}{N} \sum_{i=1}^{N} \sum_{t=t_0}^{t_\tau} \sqrt{(I_{DREC}(i,t) - I_D(i,t))^2}. \tag{8}$$

Finally, we also determined the SNR of the $I_{SPC}$ signal, which is used for the image retrieval $m$ in delay $t = 0$ µs. Note that the $I_{SPC}$ signal indicates the fluctuation of $I_{DA}$ curves corresponding to the respective delay. The fluctuations ($I_{SPC}$) are affected by the noise in both $I_{EXC}$ and $I_{PL}$. We extract the $I_{SPC}$ noise level to provide a comparison to other SPC experiments. *SPC-SNR$_{EXC}$* denotes the noise level in $I_{SPC}$ when the noise was added to $I_{EXC}$ and *SPC-SNR$_{PL}$* indicates the case when the noise was added to the $I_{PL}$. Analogously, the indices "PL "and "EXC "have the same meanings for $R_{EXC}$ and $R_{PL}$ or $\sigma_{EXC}$ and $\sigma_{PL}$.

Table 1: Quality of the retrieved image (*R*, lower number = higher quality) and SPC signal (SPC-SNR, higher number = higher quality) corresponding to Fig. 3 for noise introduced via $I_{EXC}$ and $I_{PL}$ signal.

|  | noise source | k | noise level | | | |
|---|---|---|---|---|---|---|
|  |  |  | SNR 15.2dB | SNR 18.2dB | SNR 20dB | SNR 23dB |
| R | EXC | 0.4 | 0.177 | 0.158 | 0.150 | 0.142 |
|  |  | 0.6 | 0.175 | 0.155 | 0.149 | 0.142 |
|  |  | 0.8 | 0.175 | 0.155 | 0.149 | 0.142 |
|  | PL | 0.4 | 0.328 | 0.265 | 0.242 | 0.191 |
|  |  | 0.6 | 0.312 | 0.264 | 0.215 | 0.192 |
|  |  | 0.8 | 0.326 | 0.261 | 0.224 | 0.181 |
| SPC-SNR | EXC | 0.4 | 38.01 | 43.80 | 48.14 | 54.97 |
|  |  | 0.6 | 38.06 | 43.84 | 48.05 | 55.04 |
|  |  | 0.8 | 37.98 | 43.90 | 48.25 | 55.16 |
|  | PL | 0.4 | 34.67 | 37.52 | 39.53 | 43.24 |
|  |  | 0.6 | 34.86 | 37.45 | 40.01 | 43.10 |
|  |  | 0.8 | 34.84 | 37.83 | 39.48 | 42.72 |

## 4. Noise level optimization

In Section 3, we studied the effect of the noise level in the $I_{EXC}/I_{PL}$ simulated dataset on the retrieval of the PL decay curve and PL snapshot. The noise levels are determined by the properties of an optical setup, which features a certain photon budget (number of detectable photons), types of detectors, and samples. For an optimized optical setup, such characteristics can not be easily improved.

On the other hand, it is possible to enhance the quality of the retrieved PL decay or FLIM on the expenses of the acquisition time. Therefore, we studied the benefits connected to a longer acquisition time. Due to the fact that the RATS method is based on signal deconvolution, an important factor is the periodicity of the signal, as we discuss in the following subsection.

### *4.1 Periodic extension of acquisition time*

A straightforward approach to improve the quality of the PL decay $I_D$ retrieval via Eq.(2) is to repeat the same measurement for the same excitation waveform. Hence, we attain a periodic $I_{EXC}$ and $I_{PL}$ signal. However, it is worth stressing that the use of a periodic signal $I_{EXC}$ will create unwanted artefacts in the retrieved data. It follows from the nature of deconvolution in Eq.(2) that such periodic excitation waveform leads to a periodic $I_D$ signal with a lower amplitude. We will demonstrate this effect on a 0D-RATS simulation.

We compare in Fig. 4 the entire retrieved PL decay ($I_D$ curve) in the case of a non-periodic $I_{EXC}$ signal with a length of 0.1 s (left-hand side) and a periodic $I_{EXC}$ signal (7 periods) with the same total duration of 0.1 s. Their comparison shows that the amplitude of the retrieved PL for the periodic $I_{EXC}$ signal is about seven times smaller compared to the non-periodic case and the PL signal has multiple replicas.

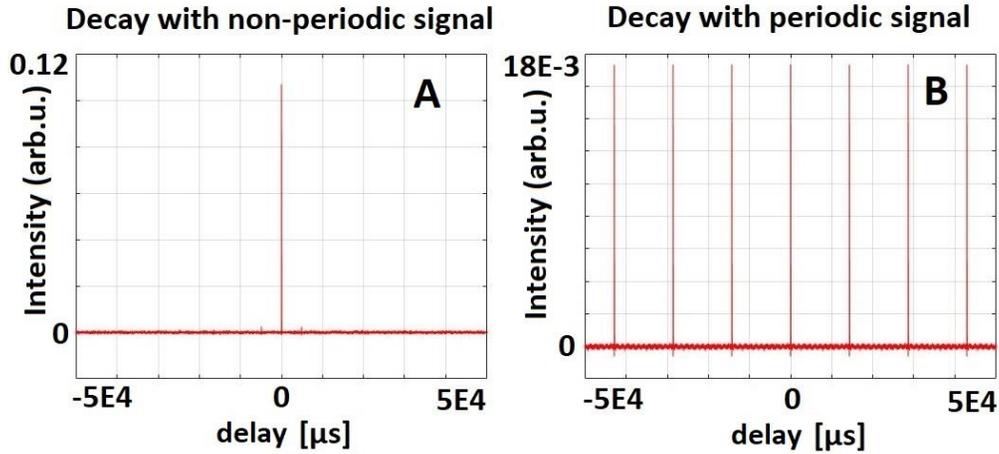

Figure 4: Results of $I_D$ reconstruction (deconvolution), when non-periodic (panel A) and periodic (panel B) $I_{EXC}$ was used for sample excitation with lifetime $\tau = 20$ μs.

When we zoom in Fig. 4(B) into the PL dynamics during the first hundreds of microseconds, i.e., one of the PL decay replicas, we get after normalization the correct $I_D$ following the actual decay – see Fig. 5. However, the resulting *SNR* is decreased. We tested in Fig. 5 the noise introduced via the excitation signal $I_{EXC}$ (panel A) and PL signal $I_{PL}$ (panel B). The root-mean-square error for curves in Fig. 5 reaches $RMSE_{PL} = 29,7.10^{-3}$ and $RMSE_{EXC} = 7,8.10^{-3}$ for the noise introduced in the $I_{PL}$ and $I_{EXC}$ signal, respectively. The noise level in Fig. 5 can be directly compared to Fig. 2, where we used a non-periodic signal and where $RMSE_{PL} = 12,9.10^{-3}$ and $RMSE_{EXC} = 2,8.10^{-3}$. For the non-periodic signal, the resulting noise level was more than 2-times lower.

To further illustrate this behavior, we carried out a set of simulations, where we kept a constant total acquisition time of 0.1 s, while the $I_{EXC}$ signal was set to be periodic with up to 16 periods – see Fig. 6. The root-mean-square error increases with the number of periods for both situations – noise in the $I_{EXC}$ and $I_{PL}$ signal. Therefore, for the sake of optimum signal reconstruction in the RATS method, it is worth avoiding any periodicity in the $I_{EXC}$ signal.

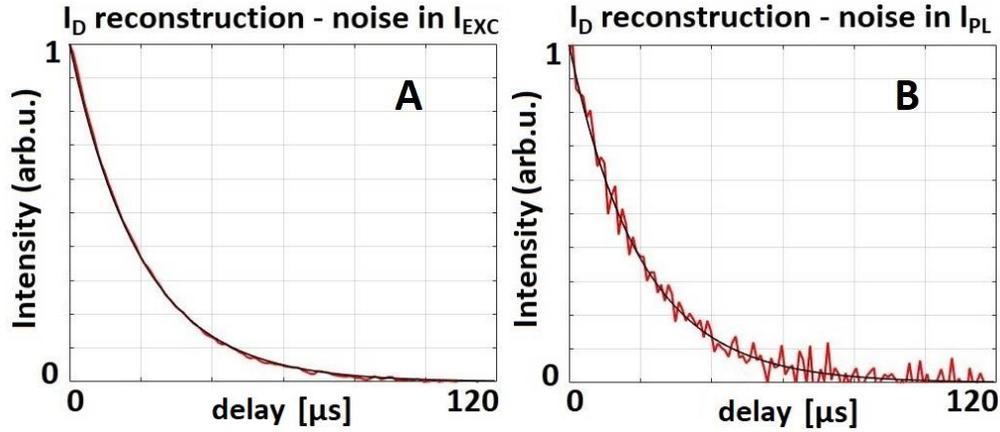

Figure 5: Zoomed results of $I_D$ reconstruction ($\tau = 20$ µs.), when $I_{EXC}$ was used as a periodic signal (7 periods). Redline is reconstructed data, and the black line is reference data. (A) Reconstruction with an amount of noise of 3% in $I_{EXC}$ (correspond with SNR about 15.2 dB). (B) Reconstruction with an amount of noise of 3% in $I_{PL}$ (correspond with SNR about 15.2 dB).

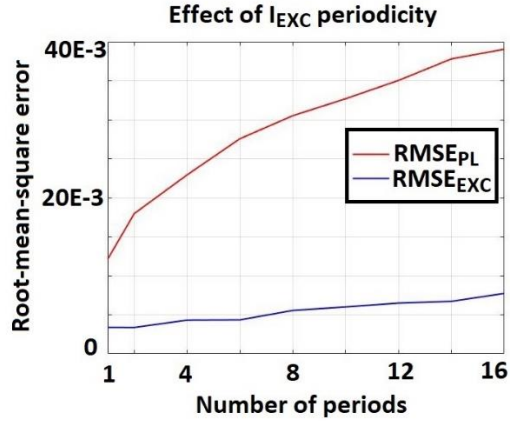

Figure 6: RMSE of $I_D$ for a periodical $I_{EXC}$ with a constant total acquisition time of $t_{acq} = 0.1$ s. The number of periods in the $I_{EXC}$ signal is varied. $I_{PL}/I_{EXC}$ noise level: SNR 15.2 dB.

### 4.2 Non-periodic acquisition time extension

The possible way to eliminate the noise effect is to extend the acquisition time so that we avoid any $I_{EXC}$ signal periodicity. This approach favors frequencies representing the true signal in the Fourier spectrum and suppresses the contribution of white noise. We carried out simulations corresponding to the 2D-RATS measurement presented in Fig. 3. Nevertheless, here we used a set of acquisition times $t_{acq} = 0.1$ s, $t_{acq} = 0.2$ s, $t_{acq} = 0.4$ s, $t_{acq} = 1$ s and $t_{acq} = 2$ s. Other conditions, such as reconstruction parameters, regularization parameter $\varepsilon$ and signal properties, were kept the same as in Section 3.

As in Section 3.2, we focus on the quality of the retrieved image ($R_{EXC}$, $R_{PL}$), noise introduced into the SPC signal ($SPC\text{-}SNR_{EXC}$, $SPC\text{-}SNR_{PL}$) in delay t = 0 µs. Moreover, this section also focuses on the deviation of the whole reconstructed PL decay in each pixel of the sample ($\sigma_{EXC}$, $\sigma_{PL}$).

Dependences of each noise characteristics on the acquisition time are presented in Fig. 7-9. In these figure, the line/symbol color indicates the given *SNR (*red: 15.2dB, blue: 18.2dB, black: 20dB and magenta: 23dB); line/symbol type represent the compression ratio $k$ (cross: 0.8, circle: 0.6, full line: 0.4).

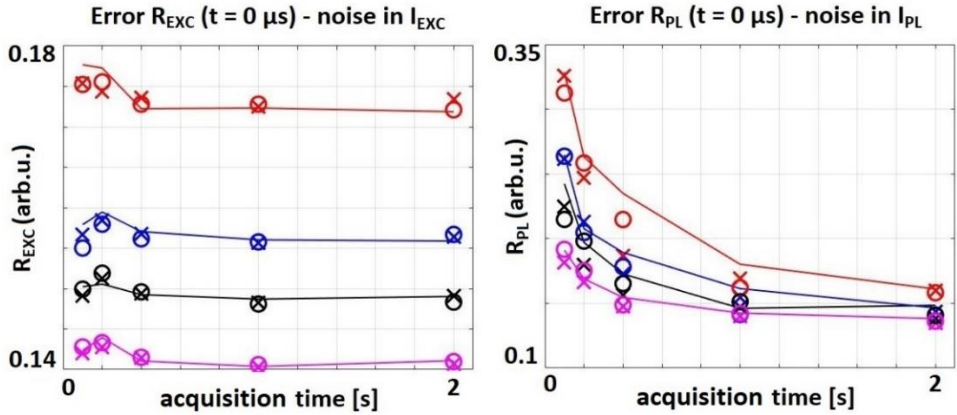

Figure 7: Reconstruction error $R_{EXC}$ (left part), $R_{PL}$ (rigth part) at the delay $t = 0$ μs dependence on sensing time for different SNR: 15.2 dB (red), 18.2 dB (blue), 20 dB (black), 23 dB (magenta) and three different compression reatio k = 0.4 (full line), k = 0.6 (circle), k = 0.8 (cross).

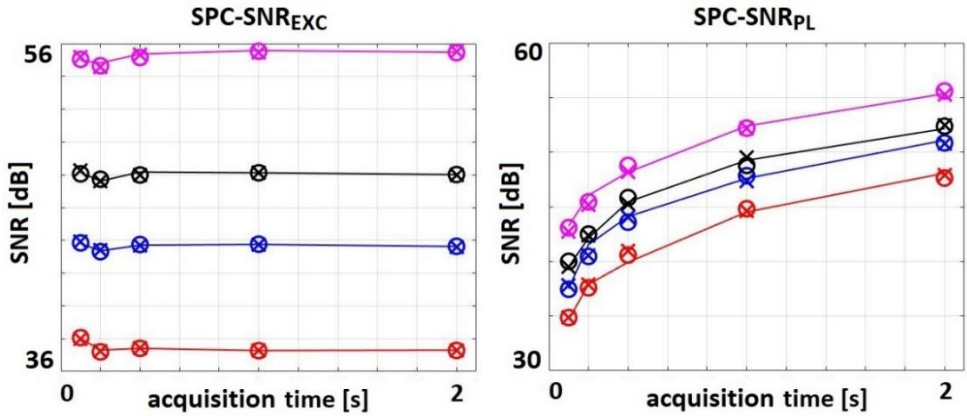

Figure 8: Noise dependence in SPC signal (random fluctuation in intensity of $I_D$ in delay $t = 0$ μs, because of changing illumination masks) on sensing time according to added noise to $I_{EXC}$ (left part) and $I_{PL}$ (rigth part) with SNR: 15.2 dB (red), 18.2 dB (blue), 20 dB (black), 23 dB (magenta) and three different compression ratio k = 0.4 (full line), k = 0.6 (circle), k = 0.8 (cross).

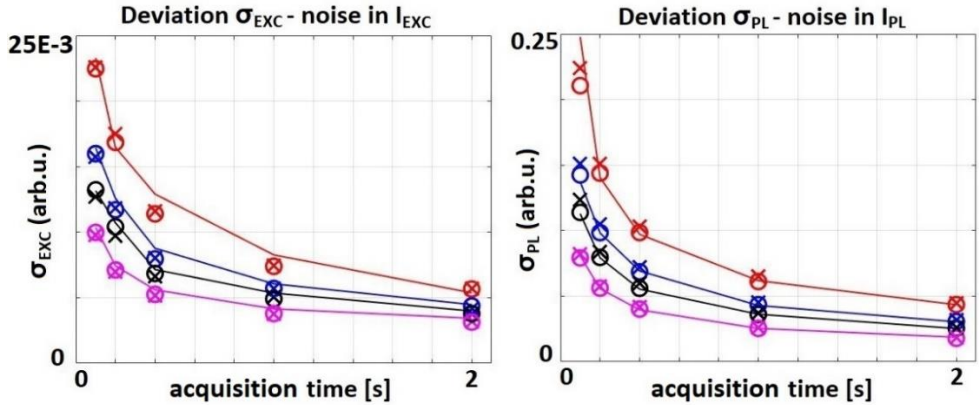

Figure 9: Acquisition time dependance of $\sigma_{EXC}$ – noise added to $I_{EXC}$ (left part) and $\sigma_{PL}$ – nosie added to $I_{PL}$ (right part). SNR: 15.2 dB (red), 18.2 dB (blue), 20 dB (black), 23 dB (magenta) and three different compression reatio k = 0.4 (full line), k = 0.6 (circle), k = 0.8 (cross).

All the results in Fig. 7-9 confirm that the effect of noise level (varying color) is much more pronounced than the compression ratio (varying line/symbol type), i.e., increasing the number of excitation patterns above 40% compared to the number of the pixel has a negligible effect on the image quality.

When we focus on the noise introduced via $I_{EXC}$ signal, it has a low effect on the resulting image reconstruction – see Fig. 3, left-hand side. For this reason, the change in the acquisition time has almost no effect on the resulting noise level regarding the image quality and SPC signal error (see Fig. 7-8).

On the contrary, for the noise introduced via the $I_{PL}$ signal, we can highly improve the quality of the retrieved image by the increased acquisition time. This image quality enhancement is much more pronounced for higher noise levels. In the case where the noise level is 15.2 dB (3%), the $R_{PL}$ will decrease by 50%, while when the noise level is 23 dB (0.5%), the $R_{PL}$ will decrease by only 25%.

The results described in Fig. 7 are linked to the results presented in Fig. 8. As expected, the highest $SPC\text{-}SNR_{EXC}$ and $SPC\text{-}SNR_{PL}$ are reached for the lowest noise level in the system - $SNR$ = 23 dB (0.5%). On the other hand, the $SPC\text{-}SNR_{PL}$ increases as the acquisition time increases, which results in a decrease in the $R_{PL}$. An interesting finding is that when the acquisition time is extended to 2s, the $SPC\text{-}SNR_{PL}$ increases even above the $SPC\text{-}SNR_{EXC}$ value corresponding to this acquisition time. Consequently, the same trend is observed for the $R_{PL}$ and $R_{EXC}$ values at a scanning time of 2 s.

Although it was stated that $R_{EXC}$ and $SPC\text{-}SNR_{EXC}$ do not change significantly with the acquisition time, it is worth noting that these values represent only a single PL snapshot at the selected delay $t = 0$ µs. The effect of acquisition time is notable for the decay curve reconstruction, which is highly improved both for the noise in $I_{EXC}$ and $I_{PL}$ -- see $\sigma$ in Fig. 9. Decay deviation $\sigma$ decreases with the increasing acquisition time. Although both errors are in a different order of magnitude, by increasing the acquisition time from 0.1 s to 2 s, both $\sigma$ errors decreased by the same ratio.

*4.3. Regularization parameter effect*

Another way to eliminate the effect of noise is to change the regularization parameter $\varepsilon$ in Eq.(2) and Eq.(4). The regularization parameter makes it possible to solve the ill-conditioned problem, where "division by zero" could occur for frequencies with very low amplitude in the $I_{EXC}$ signal [23]. The regularization parameter adds a part of the power of the spectrum to the denominator of the Eq.(2) respective Eq.(4) and thus eliminates the influence of less frequent frequencies in the signal (white noise). As a result, the calculated $I_D$ or $I_{DA}$ is smoothed.

As in the previous section, the influence of the parameter $\varepsilon$ was investigated regarding the quality of the reconstructed PL snapshot at the zero delay after excitation, which corresponds to the maximum PL intensity. All simulations were carried with the acquisition time of 0.1 s and a range of $\varepsilon$ from 0.05 to 1. Reconstruction parameters and signal properties were held the same as in Section 3. Fig. 10-12 follow the same color scheme as in the previous section with respect to $SNR$ (color) and compression ratio (line/symbol type).

In accordance with the previous results, the effect of the added noise is much more pronounced than the compression ratio used. We will first focus on the influence of regularization parameter $\varepsilon$ on the quality of the retrieved image, i.e., $R_{EXC}$ and $R_{PL}$ – see Fig 10. In these results, we see an interplay of two effects: the smoothing of the $I_{DA}$ and $I_D$, respectively, and the bias of the reconstructed $I_{DA}$ respective $I_D$. Since the noise level in the $I_{EXC}$ affects the $I_{DA}$ reconstruction less than in $I_{PL}$, the effect of biased reconstruction of $I_{DA}$ prevails and the $R_{EXC}$ error increase with $\varepsilon$. Whereas in the case of $R_{PL}$, the noise smoothing effect prevails in $I_{DA}$ for the low $\varepsilon$ values and the $R_{PL}$ decreases. However, with increasing $\varepsilon$, the bias effect becomes dominant and the image quality is decreased again.

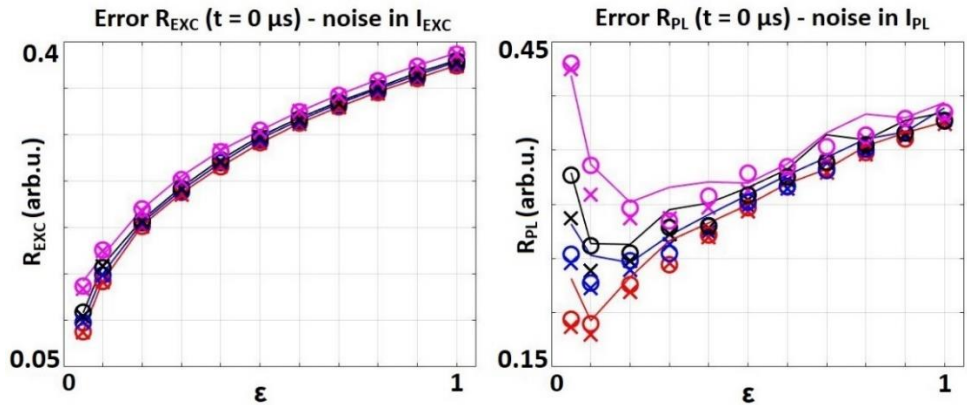

Figure 10: Reconstruction error $R_{EXC}$ (left part), $R_{PL}$ (rigth part) at the delay $t = 0$ µs dependence on ε - different SNR: 15.2 dB (red), 18.2 dB (blue), 20 dB (black), 23 dB (magenta) and three different compression reatio $k = 0.4$ (full line), $k = 0.6$ (circle), $k = 0.8$ (cross).

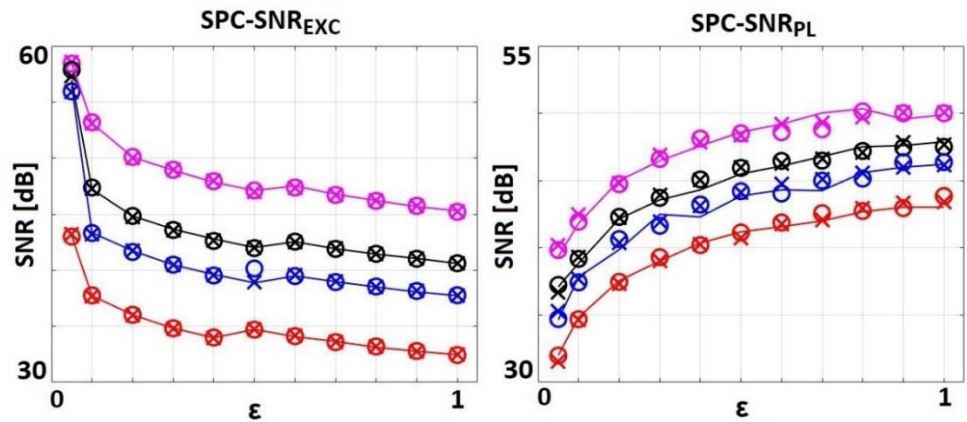

Figure 11: Noise dependence in SPC signal (random fluctuation in intensity of $I_D$ in delay $t = 0$ µs, because of changing illumination masks) on ε according to added noise to $I_{EXC}$ (left part) and $I_{PL}$ (rigth part) with SNR: 15.2 dB (red), 18.2 dB (blue), 20 dB (black), 23 dB (magenta) and three different compression reatio $k = 0.4$ (full line), $k = 0.6$ (circle), $k = 0.8$ (cross).

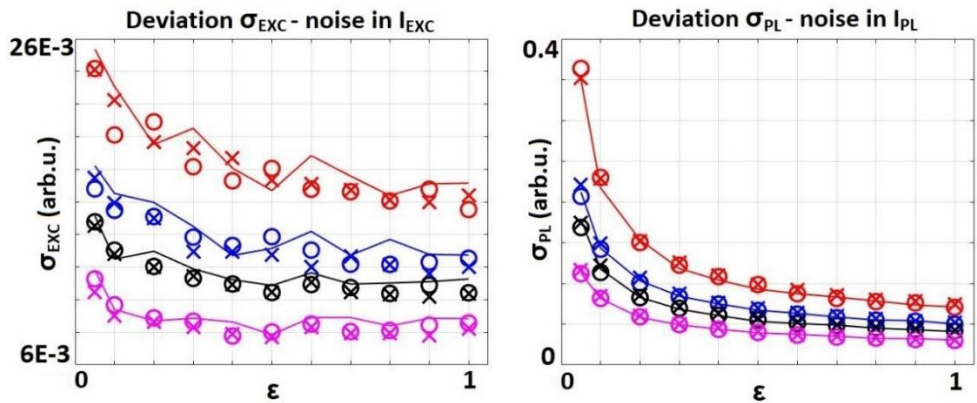

Figure 12: ε dependance of $\sigma_{EXC}$ – noise added to $I_{EXC}$ (left part) and $\sigma_{PL}$ – nosie added to $I_{PL}$ (right part). SNR: 15.2 dB (red), 18.2 dB (blue), 20 dB (black), 23 dB (magenta) and three different compression reatio $k = 0.4$ (full line), $k = 0.6$ (circle), $k = 0.8$ (cross).

This is confirmed by Fig. 11, which shows the amount of noise in the individual *SPC-SNR$_{EXC}$* and *SPC-SNR$_{PL}$*. This implies that for the given noise level in the $I_{PL}$ signal, there is an optimum regularization parameter providing the best results and this provides a guideline of the optimum $\varepsilon$ value.

Fig. 12 focuses on $\sigma$, which evaluates the fidelity of the reconstructed decay curves. Here, with increasing $\varepsilon$, both $\sigma_{EXC}$ and $\sigma_{PL}$ decrease despite the bias effect, i.e., a higher $\varepsilon$ value provides a better estimate of the decay curve. We ascribe it to the fact that with decreasing intensity of $I_{DA}$ (the following delay after excitation), the noise level in the SPC signal increases, and thus the smoothing effect of the regularization parameter prevails over the bias effect.

Both the dependence of $\sigma_{EXC}$ and $\sigma_{PL}$ on $\varepsilon$ are influenced by the amount of noise in the system. Especially for the high-noise PL signal (15.2 dB, 3%), the increase of $\varepsilon$ from 0.05 to 1 reduced the $\sigma_{PL}$ by 79%. Analogously to the previous results, the increase in regularization parameter has a lower effect on the noise introduced via $I_{EXC}$ signal -- see $\sigma_{EXC}$, which decreases by 32% for the same noise level and $\varepsilon$ change.

## 5. Conclusion

In this paper, we carried out an extensive analysis of the effect of noise within the RATS technique – a novel method for the PL decay reconstruction and FLIM. Since this is a new method, the findings presented are important for the further direction of the method. The problem was investigated using simulations for cases with noise levels reaching 0.5 -- 3% (*SNR* 23 -- 15.2dB) of the signal. To disentangle the effect of each noise source, the noise was added to $I_{EXC}$ only and $I_{PL}$ was left noiseless and vice versa.

We consistently observed that the same noise level has a significantly greater effect when present in the $I_{PL}$, which we ascribe to the fact that the regularization parameter $\varepsilon$ is associated with the $I_{EXC}$ in the denominator of the relationship Eq.(2) and Eq.(4). Hence, the most efficient strategy to improve the data quality is to focus on the $I_{PL}$ intensity level and the connected noise.

Secondly, all our data simulations revealed that the effect of noise level is incomparably more important than the compression ratio $k$, i.e., the number of measured excitation masks in the single-pixel experiment. In other words, for $k > 0.4$, it is not possible to compensate for a higher noise level by increasing the number of excitation masks.

It is also essential to avoid using periodic excitation signals, which can cause periodicity in the retrieved decay and increase the resulting $I_D$ relative noise level. It is worth noting that the RATS method was originally based on an analog random signal generator, the main component of which is a rotary diffuser. Therefore, we attained a quasi-periodic signal, which played a significant role in the method. Using a non-periodic signal, for instance, by laser modulation, a greater signal-to-noise ratio will be obtained.

As expected, we observed that increasing the acquisition time of a single decay measurement (single excitation mask) can highly reduce the resulting noise and effectively compensate for the noise present in the PL measurement. The choice of acquisition time is also connected with the *IRF* of the signal and the expected PL lifetimes. Moreover, it fundamentally manages total acquisition time. Here we used lifetime $\tau = 20$ μs, so that $I_{EXC}$ signal with *IRF* = 2.07μs was selected with data sampling 0.99μs. For other signal parameters, different results than those presented here can be achieved after the same signal length extension. If we focus on shorter lifetimes, a faster signal (smaller *IRF*) is needed, and the total scanning time is also reduced in a given ratio. However, the simulated trends for *R, SPC-SNR*, and $\sigma$ can be applied without compromising generality.

In order to eliminate signal noise, we have deliberately avoided any mathematical signal filtering techniques so that we do not lose important information due to the random nature of the signal. Mathematical signal filtering could be optimized on the expenses of the general applicability of the results. Therefore, we explored instead the possibility to optimize the regularization parameter $\varepsilon$ used in the PL decay retrieval.

The parameter $\varepsilon$ was changed in the range 0.05 to 1. We observed a trade-off between the two effects. On the one hand, a higher value of $\varepsilon$ smoothes the $I_{DA}$ curve, while, on the other hand, it distorts the $I_{DA}$ because some frequencies of the Fourier spectrum are favored. Therefore, we observed that there is an optimum value of $\varepsilon$ connected to the noise level in the $I_{PL}$ signal. As a rule of thumb, for a system with a low noise level, we recommend keeping the parameter ε at 0.1 or 0.2. For a system with a high noise level, it is possible to increase the regularization parameter more significantly.

**Funding**

Ministry of Education, Youth and Sports ("Partnership for Excellence in Superprecise Optics," Reg. No. CZ.02.1.01/0.0/0.0/16_026/0008390) and the Student Grant Competition at the Technical University of Liberec (under the projects No. SGS-2021-3003).

**Acknowledgment**

We gratefully acknowledge the Mechanics laboratory of TOPTEC (Institute of Plasma Physics, Czech Academy of Sciences) for providing us with unique optomechanics components since the beginning of RATS method development.

**Disclosure**

The authors declare no conflicts of interest.